\documentclass[manuscript]{aastex}
\def\lesssim{\mathrel{\hbox{\rlap{\hbox{\lower4pt\hbox{$\sim$}}}\hbox{$<$}}}}
\def\gtrsim{\mathrel{\hbox{\rlap{\hbox{\lower4pt\hbox{$\sim$}}}\hbox{$>$}}}}

\def\ltsima{$\;\buildrel < \over \sim \;$}
\def\simlt{\lower.5ex \hbox{\ltsima}}
\def\gtsima{$\;\buildrel > \over \sim \;$}
\def\simgt{\lower.5ex \hbox{\gtsima}}
\def\lesssim{\mathrel{\hbox{\rlap{\hbox{\lower4pt\hbox{$\sim$}}}\hbox{$<$}}}}
\def\gtrsim{\mathrel{\hbox{\rlap{\hbox{\lower4pt\hbox{$\sim$}}}\hbox{$>$}}}}
\def\gtrless{\mathrel{\hbox{\rlap{\hbox{\lower4pt\hbox{$<$}}}\hbox{$>$}}}}
\def\rightleftharpoons{\mathrel{\hbox{\rlap{\hbox{\raise2pt\hbox{$\rightharpoonup$}}}\hbox{$\leftharpoondown$}}}}
\def\notrightleftharpoons{\mathrel{\hbox{\rlap{\hbox{\raise1.5pt\hbox{$\;\mid$}}}\hbox{$\rightleftharpoons$}}}}
\def\dbar{\mathrel{\hbox{\rlap{\hbox{\raise3pt\hbox{$-$}}}\hbox{$d$}}}}
\def\hbar{\mathrel{\hbox{\rlap{\hbox{\raise3pt\hbox{$-$}}}\hbox{$h$}}}}
\def\nubar{\mathrel{\hbox{\rlap{\hbox{\raise2pt\hbox{$-$}}}\hbox{$\nu$}}}}
\def\lambdabar{\mathrel{\hbox{\rlap{\hbox{\raise2pt\hbox{$-$}}}\hbox{$\lambda$}}}}
\def\BbbV{\mathrel{\hbox{\rlap{\hbox{\raise2.5pt\hbox{${\rm v}$}}}\hbox{${\rm V}$}}}}
\def\BbbT{\mathrel{\hbox{\rlap{\hbox{\raise2pt\hbox{${\rm T}$}}}\hbox{${\rm T}$}}}}

\def\dddot{\hbox{\rlap{\hbox{\raise 8pt\hbox{${\bf ...}$}}}\hbox{$$}}}

\def\ltsima{$\;\buildrel < \over \sim \;$}
\def\simlt{\lower.5ex \hbox{\ltsima}}
\def\gtsima{$\;\buildrel > \over \sim \;$}
\def\simgt{\lower.5ex \hbox{\gtsima}}

\shorttitle{Self-Sustaining Star Formation} 
\shortauthors{Harwit}

\begin{abstract}

The decline of star formation in massive low-redshift galaxies, often referred to as ``quenching," has been attributed to a variety of factors.  Some proposals suggest that erupting active galactic nuclei may strip galaxies of their interstellar medium, ISM, and thus the ability to form stars.  Here we note that, whereas star formation is universal in small, low-redshift galaxies, fractional duty cycles $\Phi$ of star-formation steadily decline in galaxies of increasing mass, although star formation may not cease entirely.  We show that, when infall of gas from extragalactic space ceases, galaxies of high stellar mass $M_*$ appear to sustain star formation on gas liberated in mass loss from evolved low- and intermediate-mass stars admixed with occasional Type II supernova ejecta.  This model quantitatively accounts for the universal limiting metallicity plateau at a ratio of oxygen to hydrogen atoms, $Z_O = n_O/n_H \sim 1.3\times 10^{-3}$, characterizing high-mass intermittently star-forming galaxies. We show that, when duty cycles $\Phi$ are specifically taken into account, the star formation rates of galaxies on this plateau correspond to mass loss rates from evolving stars of order $\dot M\sim 10^{-11}M_*$ yr$^{-1}$, in rough agreement with observed estimates. Far-infrared continuum and fine-structure line observations, as well as molecular data, may soon be able to resolve whether or not low levels of sporadic star formation can be sustained indefinitely in massive galaxies.

{\it Key words:} galaxies: abundances --- galaxies: evolution --- galaxies: star formation ---ISM: abundances --- stars: abundances
\end{abstract}

\begin{document}

\title{{\bf Intermittent Self-Sustaining Star Formation in Massive Low-Redshift Galaxies, Exhibiting a Peak Metallicity Plateau}}
\author{Martin Harwit\altaffilmark{1,2}}

\email{harwit@verizon.net}

\altaffiltext{1}{Center for Radiophysics and Space Research, Cornell University, Ithaca, NY, 14853}
\altaffiltext{2}{511 H street, SW, Washington, DC 20024-2725}

\section{INTRODUCTION}\label{Section:Introduction}

A variety of processes have recently been proposed to account for the apparent cessation of star formation in massive, red galaxies.  Simulations by Springel et al., (2005) suggested that AGN outbursts could strip gas from a galaxy's inner regions to quench star formation.  Simulations by Bekki (2009) suggest that ram-pressure stripping of a galaxy's halo gas by the intergalactic medium could similarly stop star formation.  McCarthy et al., (2011) examined the extent to which AGN outflows could control star formation in groups of galaxies, and Ishibashi and Fabian (2015) analyzed radiation-pressure-driven AGN outflows that might curtail star formation in massive galaxies.  

Although Fischer et al., (2010) showed that AGN outflows can massively eject a host galaxy's interstellar medium, ISM, it is unlikely that quenching of star formation can generally be attributed to such events.  Most quenched galaxies do not exhibit signs of a violent past, nor do many of them appear to harbor active galactic nuclei, at least as indicated by the Sloan Digital Sky Survey, SDSS, which shows an abundance of high-mass galaxies devoid of either AGNs or measurable SFRs.  Criteria specified by Kauffmann et al. (2003) can reliably identify and remove AGNs from lists of SDSS galaxies.

Moreover, even if an AGN were to sweep a galaxy's entire ISM out of a galaxy, mass-loss from evolved stars would soon replenish the interstellar gases, even in the absence of extragalactic infall, so that a permanent cessation of star formation would still need to be explained.  One would rather expect, instead, that the ISM of galaxies steadily accumulating gases ejected by evolved low- and intermediate- mass stars --- hereafter referred to simply as ``mass loss" --- would be subject to punctuated episodes of star formation exhibiting duty cycles, $\Phi\lesssim 1$.  An examination of what happens when all infall ceases is thus of particular interest.  Insight on this can be gained from the metallicities of SDSS galaxies.

\begin{figure}[t!]
\centering
\includegraphics[height=0.85\textwidth,angle=-90, trim= -.75cm -1.25cm -2.75cm -5.05cm, clip=true]{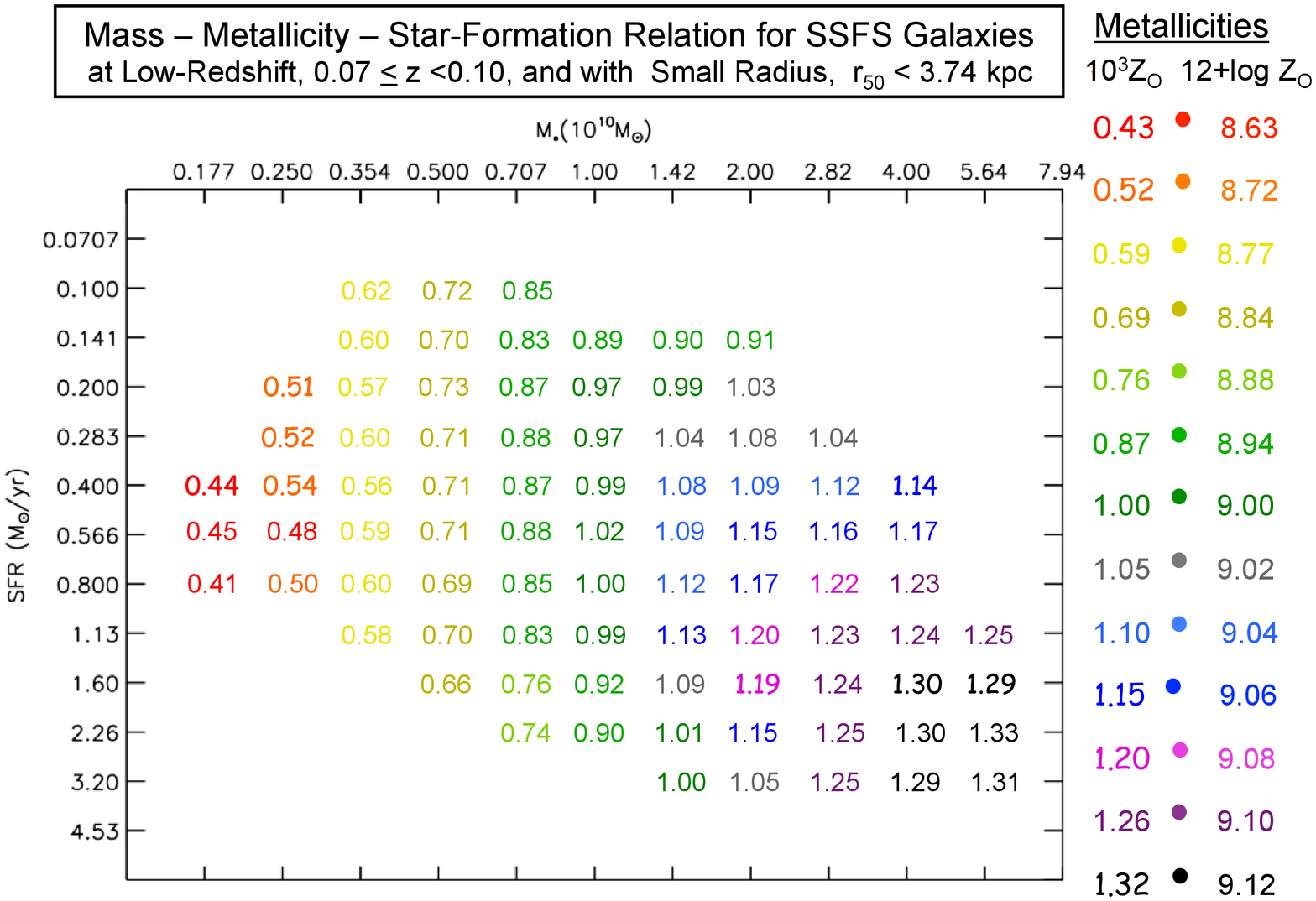}
\caption{Metallicities of small-radius, low-redshift, Strongly Star Forming Sources, SSFS, for SFR/$M_*$ bins containing $\geq 50$ galaxies.  The narrow range of radii $< 3.74$ kpc and redshifts $0.07\leq z\leq0.10$ emphasizes features partially lost when metallicities are averaged over the broader ranges covered by Mannucci et al. (2010).  On the right, metallicities $Z_O$ given as number densities of oxygen relative to hydrogen atoms, and displayed as $10^3Z_O$ in the table, are translated into their widely used logarithmic values, $12+\log Z_O$.  Metallicity tables for other radius and redshift ranges are available online, appended to Paper 2.}
\label{fig:MZSFR}
\end{figure}

The SDSS, provides wide-ranging data on oxygen metallicities $Z_O$ of galactic H$_{\rm II}$ regions, their star formation rates, SFR, and their relation to galaxy stellar mass, $M_*$.  (Tremonti et al., (2004), Nagao et al., (2006),  Mannucci et al., (2010), Lara-Lopez et al., (2010), Zahid et al. (2014)). Oxygen is a useful tracer of star formation rates because Type II SNe explosions are the main source of cosmic oxygen and thus become gauges of star formation rates once an IMF is established and reliable models of stellar evolution for stars of different masses are  in hand. 

One feature of the mass/metallicity/star-formation relation based on oxygen abundances is that quenching of star formation at first appears paradoxical:   Figure \ref{fig:MZSFR} shows metallicities of H$_{\rm II}$ regions in low-mass SDSS galaxies having specific star formation rates, SSFRs $\lesssim 2.5\times 10^{-10} M_{\odot}$ per year.  Because of a high rate of extragalactic infall, the metallicities of these galaxies, determined by ratios of oxygen to hydrogen by number density of atoms $n_O/n_H\equiv Z_O$, are as low as $Z_O\sim 0.45\times 10^{-3}$.  Metallicities then rise steadily with increasing $M_*$ toward a plateau at masses $M_*\gtrsim 5\times 10^{10} M_{\odot}$, where $Z_O\sim1.3\pm 0.3\times 10^{-3}$ and SSFRs have dropped to $\sim 5\times 10^{-11} M_{\odot}$ per year.   In contrast, one might have expected a galaxy approaching a dearth in star formation due to diminishing extragalactic infall to exhibit an ISM mainly fed by mass loss from low- and intermediate-mass evolved stars.   Oppenheimer and Dav\'e (2008) --- hereafter referred to as O\&D(2008) --- showed that the mass loss from these stars would have a metallicity $Z_O\sim 0.26\times 10^{-3}$, i.e., five times lower than high-mass SDSS galaxies exhibit.   An aim of the present paper is to find clarity in these observations.

Over the past several years, Brisbin and Harwit have investigated the relationship between the infall of extragalactic matter into galaxies, the mixing of infalling gas and gases returned to the ISM through mass-loss, and the ejecta generated in supernova explosions (Brisbin and Harwit (2012), hereafter referred to as Paper 1).   Paper 1 took a first step to establish a link between dynamic and chemical processes, with the metallicity of the ISM serving as tracer of dynamics.   In a more detailed paper --- hereafter referred to as Paper 2 ---  Harwit and Brisbin (2015) documented a gradual transition in SDSS galaxies that casts serious doubt on the need for violent processes to explain quenching:  Whereas low-redshift galaxies with stellar masses $M_* \lesssim 0.5 \times 10^{10}M_{\odot}$ form stars virtually without interruption, galaxies with $M_* \gtrsim 10^{10} M_{\odot}$ progressively form stars more sporadically. 

These findings emerged from an analysis of SDSS galaxies in the redshift range $0.07 < z < 0.3$ whose emission lines were sufficiently informative to permit rough characterization of the galaxies' properties.  The set deliberately excluded galaxies displaying a dominant active galactic nucleus, AGN.  Galaxies meeting these criteria were classified as {\it Valid Normal Sources}, VNS.  Among the nearly 200,000 VNS, a further set of 82,000 galaxies were clearly forming stars according to criteria defined by the strengths of their H-$\alpha$, H-$\beta$, [NII] 6584 and [OIII] 5007 line emissions (Tremonti et al., (2004), Nagao et al., (2006)).  These galaxies were designated as {\it Strongly Star Forming Sources}, SSFS.  The advantage of dealing with SSFS was that their metallicities and SFRs were sufficiently clear that reliable mass/metallicity/SFR relations could be displayed in tabular form, as in Figure \ref{fig:MZSFR}.   Other VNS might meet some of the SSFS criteria, though not at levels sufficiently informative to establish reliable star formation rates or metallicities required for a comparable tabular display.

Paper 2 divided the SDSS data into three redshift ranges $0.07\leq z< 0.10$, $0.10\leq z <0.15$, and $0.15\leq z <0.3$, as well as three sets of Petrosian half-light radii $r_{50} < 3.74$, $3.74\leq r_{50}< 5.01$ and $r_{50}>5.01$ kpc.  These radii are determined by a 3 arc second aperture on the SDSS spectroscopic fiber which limits our observations to a central $\sim 2$ kpc radius portion of a galaxy at redshift $z = 0.07$ and a 6.6 kpc radius portion at $z= 0.3$.    As Table \ref{table:SFSS/VNS-Ratios} shows, $\gtrsim 97\%$ of VNS with redshifts $0.07< z\leq 0.10$, $r_{50}\leq 3.74$ kpc, and $M_* \sim 0.25 - 0.5 \times 10^{10} M_{\odot}$ form stars continually.   At higher masses SSFS/VNS ratios monotonically decline, to $\sim 80\%$ at $M_* = 10^{10} M_{\odot}$ and $\sim 23\%$ at $M_* = 5.6\times 10^{10} M_{\odot}$.   

We take the $N_{\rm SSFS}/N_{\rm VNS}$ ratios of galaxies $M_*$ to represent their fractional star-forming duty cycles, $\Phi$, suggesting that star formation is episodic with active periods  and duty cycles declining with rising $M_*$. Whether or not permanent quenching gradually sets in with rising galaxy mass is therefore unclear, and forms the main topic of the current investigation.  Peng, Maiolino, and Cochrane (2015), for example, recently proposed that loss of extragalactic infall of matter into a galaxy must ultimately lead to cessation of star formation in a process they called {\it strangulation}.   The topic thus is controversial:  Does star formation in massive galaxies inevitably die out, or can star formation sporadically revive, even if only at widely separated intervals?

\begin{table}[t!]
\caption{Small-Radius, $r_{50}\leq 3.74$ kpc, Low-Redshift ($z = 0.07-0.10$) SSFS and VNS Populations and their Ratios $\Phi \equiv N_{\rm SSFS}/N_{\rm VNS}$ Interpreted as  Fractional Star-Forming Duty Cycles.* }
\vskip0pt\vskip-24pt\vskip0pt
\begin{center}
\begin{tabular}{lcccccccccccl} \hline\noalign{\vskip3pt}
$M_*(10^{10}M_{\odot})$&0.250&0.354&0.500&0.707&1.00&1.42&2.00&2.82&4.00&5.64&8.00&11.3\\
 \hline\noalign{\vskip1pt}
Low-z SFSS &   730   &   1367  &    2394  &    3428  &   3957 &    4014   &   3070   &   2014   &   988  &     392 & 138 & 22\\
Low-z VNS    &   745  &    1383   &   2459   &   3788   &   5016   &   6045  &    5915   &   4761  &    2966    &   1697 & 861 & 295\\
$\Phi\equiv N_{{\rm SSFS}}/N_{{\rm VNS}}$ & 0.98 &  0.99 &  0.97 &  0.90 &  0.79 &  0.66 &  0.52&  0.42&  0.33 &  0.23 & 0.16 & 0.07\\ 
 \hline\noalign{\vskip1pt} 
\end{tabular}
\end{center}
* For similar data on galaxies with other redshifts and radii see the online tables included with Paper 2.
\label{table:SFSS/VNS-Ratios}
\end{table}

Our main approach will be to ask whether massive galaxies devoid of infall may sporadically be sustaining low levels of star formation fuelled solely by accumulations of hydrogen released in mass loss and, if so, how long this process may be sustained, particularly in view of the intermittency in star formation cited above?  The question arises because, if nothing more than a critical reservoir of hydrogen were needed to permit renewed star formation in aging galaxies, mass loss should eventually accumulate to inevitably re-ignite star formation.  But because the anticipated metallicities of these gases would be so low, $Z_O \sim 2.6\times 10^{-3}$, agreement with observed metallicities of massive galaxies could only be reached if oxygen-rich ejecta from occasionally exploding Type II SNe first admixed with this oxygen-deficient mass loss.   

In Section 2 of the paper, we present a quantitative approach to star formation, with and without extragalactic infall.  Section 3 depicts how self-sustained star formation appears to arise in these galaxies.   A final Section 4 summarizes our results and conclusions.

\section{STAR FORMATION WITH AND WITHOUT INFALL}\label{Section:StarFormation}

Papers 1 and 2 showed that oxygen abundances exhibited by low-redshift galaxies can be explained by a global process, if one assumes (i) infall of extragalactic gases mildly enriched with oxygen $Z_{i,O} \equiv (n_O/n_H)|_i \sim 0.2\times 10^{-3}$ --- an abundance Lovisari et al. (2011) observed in the far reaches of galaxy clusters; (ii) a Chabrier (2003) stellar initial mass function, IMF; (iii) stellar evolution and mildly-updated metal production rates derived from O\&D(2008); and (iv) consistent with SDSS observations, a nearly constant metallicity, nearly constant ratio of gas to stellar mass, and negligibly increasing galaxy masses over times separating successive star-forming epochs at low redshifts.

We define the metallicity $Z_{\ell}$ of an element $\ell$ as its abundance by number of atoms $n_{\ell}$ relative to hydrogen, $Z_{\ell} \equiv (n_{\ell}/n_H)$.  Like O\&D(2008) we will also assume that all metals are produced and injected into the ISM in one of two ways; (i) {\it prompt return} through supernova explosions, or (ii) {\it delayed return} through stellar mass loss.  For a specific metal $\ell$,  O\&D(2008) defined a {\it prompt yield} $Y_{P,\ell}$, as the rate at which a mass $M_{P,\ell}$ of atoms $\ell$ is returned to the ISM as a fraction of the star formation rate: $Y_{P, \ell} = \dot M_{P,\ell}/{\rm SFR}$.  Correspondingly, they defined a {\it delayed yield}, $Y_{D,\ell}$, as the rate, $\dot M_{D,\ell}$, at which delayed metal is returned to the ISM as mass-fraction of the total delayed mass loss $\dot M_D$, i.e., $Y_{P, \ell}= \dot M_{D,\ell}/\dot M_D$.  A further step was to define the delayed yield as a fraction $\epsilon_D$ of the instantaneous SFR, by writing $\dot M_D\equiv \epsilon_D {\rm SFR}$.  O\&D(2008) thus defined 
\begin{equation}
Y_{P,\ell}\equiv\dot M_{P,\ell}/{\rm SFR}\quad ; \quad Y_{D,\ell}\equiv \dot M_{D,\ell}/[\epsilon_D {\rm SFR}]\ .
\label{eq:yields}
\end{equation}
Both these yields assume integration of the SFR over the entire IMF.   By number of atoms, yields can be written as
\begin{equation}
Y_{P,\ell}=\alpha_{\ell}\epsilon_PZ_{P,\ell}\quad ; \quad Y_{D,\ell} = \alpha_{\ell} Z_{D,\ell}\ ,
\label{eq:YZ}
\end{equation}
where $\epsilon_P$ is the fraction of the IMF, by mass, promptly ejected in Type II SNe. The conversion factor $\alpha_{\ell} \equiv (0.72 m_{\ell}/m_H)$ relates the abundance ratios of O\&D(2008) by mass to abundance ratios expressed as fractional number densities $Z_{\ell}$.  The numerical factor 0.72 adjusts for the cosmic ratio of hydrogen to helium.

Conservation of oxygen in the mixing of infalling extragalactic gas with a galaxy's native gases leads to:
\begin{equation}
\dot M_i(x)Z_{i,O} + \epsilon_DZ_{D,O}{\rm SFR}(x) + \epsilon_PZ_{P,O}f_{P,R}{\rm SFR}(x) = Z_O(x)\biggl[\dot M_i(x) +\epsilon_D{\rm SFR}(x) +\epsilon_Pf_{P,R}{\rm SFR}(x)\biggr] \ .
\label{eq:MetalConservation}
\end{equation} 
Here, $\dot M_i(x)$ is the extragalactic mass infall rate; its associated star formation rate is  ${\rm SFR}(x)$;  $Z_{i,O} \equiv n_{i,O}/n_{i,H}$ is the infall's oxygen metallicity; $Z_{D,O}$ is the  metallicity of the galaxy's delayed mass loss; $Z_{P,O}$ is the metallicity of Type II SNe ejecta; $f_{P,R}$ is the fraction of Type II SNe ejecta retained in, or soon falling back into the galaxy; and $Z_O(x)$ is the metallicity of the mix of the galaxy's infalling and native gases.  

Stellar metal yields deduced by O\&D(2008) on the basis of a Chabrier (2003) IMF, and nucleosynthetic stellar evolution models due to Limongi \& Chieffi (2005), could lead to adoption of numerical values  $\epsilon_P\sim 0.2$, $\epsilon_D\sim0.38$, $Z_{P,O}\sim 6.5\times 10^{-3}$, and $Z_{D,O}\sim 0.26\times 10^{-3}$ in equation (\ref{eq:MetalConservation}).  However, Paper 2 showed that inclusion of an extragalactic infall metallicity of $Z_{i,O}\neq 0$ in the expression indicated the product $\epsilon_P Z_{P,O}$ to be $\sim 10$ to 15\% too high to provide a good fit to the highest metallicities observed in high-mass SDSS galaxies.  We corrected for this by adopting a $\sim10\%$ lower value for $Z_{P,O}$, but will obtain a slightly preferred fit, in the present paper, by setting $\epsilon_P \sim0.18$ and $Z_{P,O}\sim 6.04\times 10^{-3}$ to achieve this.  Given current uncertainties on Type II SNe yields this fit seems quite reasonable.
 
The processes displayed in equation (\ref{eq:MetalConservation}) were designed to include the broad class of galaxies in which extragalactic infall plays a major role.   An important observational finding of Paper 2, already referred to in Section \ref{Section:Introduction}, is that metallicities of galaxies with identical SFRs and masses $M_*$ change only imperceptibly, between one narrow SDSS star-forming redshift range and the next.  In the notation of equation (\ref{eq:MetalConservation}), 
\begin{equation}
\dot M_i(x) +\epsilon_Pf_{P,R}(x) SFR_C(x) + \epsilon_DSFR_C(x) \sim SFR_{(C+1)}(x)\ ,
\label{eq:M/SFR}
\end{equation}
where the left side of the equation presents the amount of gaseous mass made available from extragalactic infall, supernova ejecta retained in the galaxy, and stellar mass loss, during a star-forming cycle $C$, for a next generation of stars to form in the immediately succeeding cycle ($C+1$).  Thus, setting ${\rm SFR}_C(x)\sim {\rm SFR}_{(C+1)}(x)$,
\begin{equation}
\dot M_i(x)/{\rm SFR}(x) \sim 1 - \epsilon_Pf_{P,R} - \epsilon_D\ .
\label{eq:imperceptible}
\end{equation}
The implication of this is readily apparent:  In galaxies sufficiently massive to retain all supernova ejecta, insertion of $\epsilon_P\sim 0.18$, $\epsilon_D\sim 0.38$, yields $\dot M_i(x)$/SFR$(x)\sim 0.44$.   For low-mass galaxies in Figure \ref{fig:MZSFR}, where retention $f_{P,R}\ll 1$, we obtain, instead, $\dot M_i(x)$/SFR$(x)\lesssim 0.62$.  As  a result, the ratio of extragalactic infall to star formation rate is roughly $ 0.5\pm 20\%$ across the whole mass range and for all star formation rates displayed in Figure \ref{fig:MZSFR}.  

We emphasize these points because, as we turn to galaxies devoid of infall or outflow, we wish to show that their properties logically follow from the same conservation relation (\ref{eq:MetalConservation}) by setting $\dot M_i(x) = 0$ and $f_{P,R} = 1$.  In this case stellar mass loss plus supernova ejecta become the sole sources of gas promoting star formation --- meaning that $\epsilon_D\equiv \dot M_D$/SFR, as defined above, leads to setting $\epsilon_D = (1 - \epsilon_P) = 0.82$.  Inserting these conditions into equation (\ref{eq:MetalConservation}) immediately leads, to metallicities 
\begin{equation}
Z_O(x) = \frac {\epsilon_DZ_{D,O} + f_{P,R}\epsilon_PZ_{P,O}}{\epsilon_D+f_{P,R}\epsilon_P} = \frac{(0.82\times 0.26 +0.18\times 6.04f_{P,R})\times 10^{-3}}{1} =  1.3\times 10^{-3} \quad {\rm for}\quad f_{P,R} = 1\ ,
\label{eq:plateau}
\end{equation}
showing that, for galaxies devoid of infall or outflow, self-sustaining star formation determines  a limiting metallicity plateau at $Z_O(x) = 1.3\times 10^{-3}$. 

An equally important consequence of relation (\ref{eq:plateau}) is that it holds regardless of $M_*$ or SFR.   Figure (\ref{fig:MZSFR}) substantiates this with its uniform metallicity plateau at $Z_O(x)\sim 1.3\times 10^{-3}$, ranging over a set of galaxies in high-$M_*$ columns in their highest star formation regimes. 

In Table \ref{table:plateau}, we concentrate on galaxies in the metallicity range $Z =1.30\times 10^{-3} \pm 3\%$ and compare data on small radius galaxies, respectively, at low and at medium redshifts, and in galaxy mass ranges $M_* = 4.0\times 10^{10}$ and  $5.6\times 10^{10} M_*$ covering regions on the metallicity plateau. For these we obtain mass loss rates required for sustaining star formation.  They are obtained from products of fractional mass loss $\epsilon_D$ and mean specific star-forming rates, defined as $\langle{\rm SSFR}\rangle\equiv\Phi(M_*)\Sigma (N\times$SFR)/$\Sigma NM_*$, and are in rough agreement with observed mass losses from low- and intermediate-mass stars in massive galaxies. 

Elbaz et al., (2011) found that at redshifts $z\sim 0$ galaxies on the galaxy main sequence exhibit specific star formation rates of $10^{-10} M_{\odot} \pm 50\%$ yr$^{-1}$ and, as O\&D(2008) and Segers, et al., (2015) showed, around 35-38\% of these SFRs are due to recycled stellar mass loss, yielding a mean mass-loss rate of $\sim 3.6\pm 1.8\times 10^{-11} M_{\odot}$ yr$^{-1}$ for these main sequence galaxies.  Although the plateau galaxies we are considering here are not typical main sequence galaxies, their mass losses indicated in Table \ref{table:plateau} emanate from similar types of stars and are roughy comparable. The two redshift and $M_*$ ranges considered in the Table provide data on those plateau regions for which the highest signal-to-noise ratios are available, and confirm that the SFRs averaged over these regions are roughly self-consistent.

These all are new results, although a valuable approach foreseeing the potential significance of such studies dates back to a seminal paper by Richard Larson (1972) written 40 years earlier, when far less information on the evolution of stars and galaxies was available.  To date, there has been no physical explanation for the particular metallicity value reached at the mass/metallicity/star-formation relation's limiting plateau.  Now, we have a quantitative physical relation solidly anchoring the metallicity on the plateau to accepted stellar evolution models, and the plateau's SFR$(M_*)$ to observed low- and intermediate-mass stellar outflow rates in $M_*$ galaxies.

\begin{table}[t!]
\caption{Number of SSFS Galaxies, N, per $M_*$ Galaxy and SFR Range, and the Mean Sustaining Mass-Loss Rate $\equiv \epsilon_D\langle$SSFR$\rangle\equiv \epsilon_D\Phi(M_*)\Sigma$($N\times$SFR)/$\Sigma NM_*$ on the Plateaus of Small-Radius, Low- and Medium-Redshift Galaxies of Masses $M_* = 4.00\times 10^{10}M_{\odot}$ and $5.64 \times 10^{10}M_{\odot}$.} 
\begin{center}
\begin{tabular}{lccccl} \hline\noalign{\vskip3pt}
Redshift, $z$&$0.07\rightarrow0.10$&$0.07\rightarrow0.10$& $0.10\rightarrow01.5$&$0.10\rightarrow01.5$&\\
 \hline\noalign{\vskip1pt}
SFR&$M_{*,\ell}(4.00)$&$M_{*,\ell}(5.64)$&$M_{*,m}(4.00)$&$M_{*,m}(5.64)$&($10^{10}M_{\odot}$)\\
1.60&171&55&162&64&$N(M_*,z,$\ SFR)\\
2.26&130&60&197&76&$N(M_*,z,$\ SFR)\\
3.20&74&51&157&87&$N(M_*,z,$\ SFR)\\
4.53&&&&77&$N(M_*,z,$\ SFR)\\
6.40&&&&51&$N(M_*,z,$\ SFR)\\
 \hline\noalign{\vskip1pt}
$\Phi(M_*)$&0.33&0.23&0.33&0.15&\\
$\Sigma N$&375&166&516&355&\\ 
$\Sigma$($N\times$SFR)&804&387&1206&1227&($M_{\odot}{\rm\ yr}^{-1}$)\\
$\Sigma NM_*$&1.5&0.94&2.06&2.00&($10^{13} M_{\odot}$)\\ 
Mass-Loss Rate&1.45& 0.78 & 1.58 &0.75&($10^{-11}M_*$yr$^{-1}$)\\
 \hline\noalign{\vskip1pt} 
\end{tabular}
\label{table:plateau}
\end{center}
* Entries for $\Phi(M_*)$ assume $\epsilon_D = 0.82$ and that plateau galaxies have the same $N_{\rm SSFS}/N_{\rm VNS}$ population ratios as other galaxies in their respective $M_*$ ranges.  Data for the two medium-redshift $M_{*,m}$ ranges are taken from Table 1 of Paper 2.  Tables of galaxy data for other redshift ranges and SFRs are archived with the online version of Paper 2.\\

\label{table:specific}
\end{table}

The particular value of $Z_O$ at which the plateau is observed signifies (i) that the ratio of the ejecta, respectively from massive and low mass stars, is in accord with slightly modified yields comparable to those derived by O\&D (2008) on the basis of stellar evolution models due to Limongi \& Chieffi (2005) and a Chabrier (2003) IMF; and (ii) that the mixing of these ejecta occurs on a time scale short compared to the onset of full protostellar collapse and star formation.  Such short time scales for mixing are required also by star formation regions in low-mass galaxies, whose metallicities always are higher than either the metallicity of extragalactic infall or the mass loss from low mass stars.  But it needs emphasizing in the case of massive galaxies, whose star-forming rates appear to be intermittent.  We will discuss this question further in Section \ref{Section:Self-Sustained}.

Peng et al., (2015) have depicted a {\it strangulation} of galaxies in terms of a progressive depletion of gas from galaxies devoid of infall or any other replenishment, until no gas at all remains and star formation entirely ceases.  Their approach neglects that stellar mass loss continually replenishes a galaxy's ISM.  Replenishment is a major contributor, as O\&D(2008) and, more recently, Segers et al. (2015) have emphasized.  In the low redshift universe, these authors, respectively, estimate that $\sim 38\%$ or $\sim 35\%$ of star formation regularly feeds on the mass loss from low- and intermediate-mass stars --- a significant contribution needing to be taken into account.    Equation (\ref{eq:plateau}) has done this, and appears to lead to an explanation of an observation that previously was puzzling. It thus provides essential insight on the nature of self-sustained star formation. 

\section{SELF-SUSTAINED STAR FORMATION}\label{Section:Self-Sustained}

We first return to the intermittency observed in star formation in high-$M_*$ galaxies.  This could be explained naturally by two factors.  The first is a potential loss of mass to extragalactic space through supernova explosions or AGN outbursts.  Such explosions might drag major portions of a galaxy's ISM beyond its gravitational reach.  It could then take a long time for stellar mass loss to accumulate a sufficient reservoir of interstellar gas to reignite star formation.  

However, this is not a likely cause of intermittency in a galaxy in which retention of Type II supernova ejecta is complete, as assumed in equation \ref{eq:plateau}.   Instead, we need to recall that we detect star formation in these massive galaxies solely during the potentially short periods during which the most massive stars in an IMF are emitting ionizing radiation.  Thereafter, once the most massive stars in the IMF have exploded, a prolonged period ensues during which the supernova ejecta have ample time to diffuse through their surroundings and mix with the mass loss from ambient stars.  With this mixing complete and adequate time for protostellar cooling, a next generation of star formation can then set in with ISM metallicities identical to those observed on the plateau.  To be consistent with the low-redshift SDSS data on which the observed plateau metallicities are based, the IMF in these massive galaxies would need to generate heavy elements at essentially the rate associated with a Chabrier (2003) IMF.  The fractional star-forming duty cycles, $\Phi(M_*)$, of these massive galaxies should then be given by the entries in Table \ref{table:SFSS/VNS-Ratios} for the ratios of galaxies shown to be actively forming stars at any given epoch.

We next need to ask:  Can star formation be reignited if supernovae ever succeed in sweeping all gases out of a galaxy after extant star-forming regions have already lost all their most massive stars?  Mass-loss from low-mass stars undoubtedly will continue, but observationally we lack the evidence to assert that clouds of these gases can cool, collapse, and form stars spontaneously without any influx of higher metallicity gases.  If they did, we should observe regions of star formation in which the oxygen abundances were those associate with mass loss from low-mass stars having metallicities $Z_O(x) = 0.26\times 10^{-3}$, i.e., five times lower than that on the metallicity plateau. 

We see no evidence for such low metallicities in Figure \ref{fig:MZSFR}.  Nor does a search through the table of 82,027 Strongly Star-Forming Sources, SSFS, in the on-line data stored with Paper 2 show evidence for the existence of sources with such low metallicities even among galaxies as massive as $M_* = 2.3\times 10^{11} M_{\odot}$.  The only SDSS galaxies with such low metallicities lie in the mass range $M_*\leq 1.8\times 10^9 M_{\odot}$, where star formation is virtually continuous, the ratio of extragalactic infall to local mass-loss is exceptionally high, and the low observed metallicities reflect abundances in a mix of mass loss and infall accreting onto regions of star formation that may have originally ignited at higher metallicity.  

Despite these findings, however, we cannot categorically rule out ongoing star formation in these aging, massive galaxies, even though they fail to exhibit the expected low-metallicity earmarks just discussed.  Their absence may simply be due to the way that star-formation in SDSS galaxies is usually determined, i.e, through identification of H$_{\rm II}$ regions presumed to accompany all star formation.  

This has two reasons:  

(i) Optical techniques provide rather poor tools for determining star formation rates in all but the most readily established cases.  This is because dust obscuration effects need to be carefully taken into account, and only high-signal-to-noise ratio data can be deemed reliable.  This is the reason why, out of roughly a million SDSS galaxies, only about 82,000 qualified for inclusion in our list of Strongly Star Forming Sources.   Far-infrared fine-structure-line data may ultimately yield considerably more reliable abundances of O[III], but not necessarily more reliable ratios of ionized oxygen to ionized hydrogen in star forming regions. 

(ii) The optically available data on high-mass galaxies thus only rule out significant star formation with normal IMFs.  But this may not mean that low-metallicity, low-mass stars might not be forming.  If low-metallicity gases were to form stars with IMFs favoring low-mass stars too cool to ionize their surroundings, the optical techniques the SDSS provides would not have identified them.  Far-infrared observations, however, could identify emission from low-mass stars still embedded in dust cocoons within which they might have hatched. [CII] 158 $\mu$m emission emanating from these cocoons might then signify ongoing star formation, at least in massive galaxies with an IMF biased to emphasize low mass stars.  Alternatively, such far-infrared data could clarify whether star formation has come to a complete stop there, or whether low metallicity mass loss will lead to continuing star formation after all.  

One other alternative also needs to be considered.  Conroy, van Dokkum \& Kravtsov (2015) have suggested that pervasive heating of the ISM by stellar winds or Type I SNe explosions may counter the radiative cooling of interstellar gases sufficiently to prevent star formation in massive galaxies.  At high redshifts they indicate that this might be possible.  However, their calculations, whose implications are illustrated in their Figures 2, 3, and 4, show neither type of heating adequate to this task for the physical parameters documented to date in low-redshift SDSS galaxies.   

Nevertheless, some explanation for the apparent absence of star formation in massive, red, low-redshift galaxies, is rather urgently needed.  Whether this is merely due to a failure of current observational techniques, or the absence of an appropriate theoretical explanation remains a question requiring resolution.

\section{SUMMARY AND CONCLUSIONS}

We have explored the evidence for episodic star formation ongoing in massive SDSS galaxies.  

A positive indication that this type of process may be common is the satisfactory match between observed oxygen metallicity plateaus at $Z_O\sim 1.3\times 10^{-3}$ in these galaxies and metallicities expected if mass loss from low- and intermediate-mass stars were to mix efficiently with Type II SNe ejecta in the absence of infall and outflow.  An extension of a model developed in our earlier Paper 2  leads to our equation (\ref{eq:plateau}), showing that the mix of gases with these metallicities in proportions expected from stellar evolution models should converge on this metallicity value $Z_O$.  

A second piece of evidence also is novel and addresses a long-standing puzzle of why oxygen metallicities in Figure \ref{fig:MZSFR} remain essentially constant over a plateau ranging over an extended range of galaxy masses $M_*$ and SFRs, exhibited solely at high galaxy masses $M_*$ and high SFRs.  Equation (\ref{eq:plateau}) explains this lack of significant gradients in the high $M_*$, high SFR  regime.  

Thirdly, not only do we find these massive galaxies to have the anticipated metallicities expected in episodic star formation; Table \ref{table:specific} also shows that mass loss rates required to sustain SFRs at the observed levels appear to be in place in galaxies whose mass-dependent fractional duty cycles, $\Phi(M_*)$, are inferred from their ratios of Strongly Star Forming Sources to Valid Normal Sources, $N_{\rm SFSS}/N_{\rm VNS}$.   The fallow intervals separating star-forming epochs in these galaxies, appear to be sufficiently long to provide accumulated mass losses able to sustain the observed star-forming rates. 

Equation (\ref{eq:plateau}), together with tabulated on-line data included with Paper 2,  thus suggests that low levels of self-sustained star formation may survive for gigayears in some galaxies, in mass ranges as high as $M_*\gtrsim 10^{11} M_{\odot}$, most of which exhibit SDSS-derived metallicity plateaus at $Z_O\sim 1.3\times 10^{-3}$ even though the ratios $N_{\rm SFSS}/N_{\rm VNS}$ in these online-archived SDSS samples rapidly drop to merely a few percent beyond  $M_*\geq 8\times 10^{10}M_{\odot}$.  

On the other hand, observationally we cannot yet rule out that star formation with IMFs favoring low-mass star formation fuelled by mass loss might alternatively be ongoing in high-mass galaxies. The masses of stars thus formed might simply be too low to produce the ionization levels usually associated with star formation.  Far-infrared fine- structure line or molecular spectroscopy might be required to detect or rule out such alternatives.

The present study was supported by NASA through subcontracts 1393112 and 1463766 the Jet Propulsion Laboratory awarded to Cornell University.  The author appreciates the helpful comments of Drew Brisbin, as well as the informed criticism of the referee, Benjamin D. Oppenheimer, whose thoughtful insistence on added clarity greatly improved this paper.

\section*{REFERENCES}

\noindent Bekki, K. 2009, MNRAS, 339, 2221-2230\\
Brisbin, D. \& Harwit, M. 2012, ApJ, 750, 142   [Paper 1]\\
Chabrier, G. 2003, PASP, 115, 763\\
Conroy, C., van Dokkum, P.G., \& Kravtsov, A. 2015, ApJ, 803:77\\
Elbaz, D., Dickinson, M., Hwang, H. S., et al. 2011, A\&A 533, A119\\
Fischer, J.,  Sturm, E, Gonz\'alez-Alfonso, et al. 2010, A\&A 518, L41\\
Harwit, M. \& Brisbin, D. 2015 ApJ, 800, 91   [Paper 2]\\
Ishibashi, W. \& Fabian, A. C. 2015, MNRAS, 451, 93\\
Kauffmann, G., Heckman, T. M., Tremonti, C., et al. 2003, MNRAS, 346, 1055\\
Lara-L\'opez, M. A., Cepa, J., Bongiovanni, A., et al. 2010, A\&A, 521, L53\\
Larson, R. 1972, Nature, 236, 7L\\ 
Limongi, M. \& Chieffi, A. 2005, ASP Conf. Ser. Vol. 342, ``1604 Ð 2004: ``Supernovae as Cosmological LighthousesÓ, eds. Turatto M., Benetti, S., Zampieri, L. \& Shea, W.Ó Astron. Soc. Pac., San Francisco, p 122\\
Lovisari, L. Schindler, S., \& Kapferer, W. 2011 A\&A, 528, A60\\
McCarthy, I. G., Schaye, J., Bower, R. G., et al. 2011, MNRAS, 412, 1965\\
Mannucci, F., Cresci, G., Maiolino, R., et al. 2010, MNRAS, 408, 2115\\
Nagao, T,. Maiolino, R., \& Marconi, A. 2006, A\&A, 459, 85\\
Oppenheimer, B. D. \& Dav\'e, R. 2008, MNRAS, 387, 577   [O\&D(2008)]\\
Peng, Y., Maiolino, R., \& Cochrane, R. 2015, Nature, 521, 192\\
Segers, M. C., Crain, R. A., Schaye, J. et al. 2015,  MNRAS, doi:10.1093\\
Springel, V., Di Matteo, T., \& Hernquist L. 2005, MNRAS 361, 776\\ 
Tremonti, C. A., Heckman, T. M., Kauffmann, G., et al. 2004, ApJ, 613, 898\\ 
Zahid, J. J., Dima, G. I., Kudritzki, R.-P., et al. 2014, ApJ, 791:130\\

\end{document}